\documentclass[a4paper,11pt]{article}
\usepackage{epsfig,cancel,cite,graphicx,epstopdf,amsmath,amssymb,amsfonts,slashed,dsfont,hyperref}

\newlength{\xtrawidth}
\newlength{\xtraheight}
\newcommand{\be}{\begin{equation}}
\newcommand{\ee}{\end{equation}}
\newcommand{\beq}{\begin{equation}}
\newcommand{\eeq}{\end{equation}}
\newcommand{\ba}{\begin{array}}
\newcommand{\ea}{\end{array}}
\newcommand{\bea}{\begin{eqnarray}}
\newcommand{\eea}{\end{eqnarray}}
\newcommand{\bean}{\begin{eqnarray*}}
\newcommand{\eean}{\end{eqnarray*}}

\usepackage{color}

\newcommand{\I}{{\rm i}}

\newcommand{\e}{\mathrm{e}}

\newcommand{\SU}[1]{\ensuremath{\mathrm{SU}(#1)}}

\def\d{\textrm{d}}

\def\p{\partial}

\def\fnote#1#2{\begingroup\def\thefootnote{#1}\footnote{#2}
     \addtocounter{footnote}{-1}\endgroup}

\numberwithin{equation}{section}

\begin{document}
\title{{\vskip 1cm \bf\LARGE Heterotic Calabi-Yau Compactifications with Flux\\[0.5cm]}}
\author{Michael Klaput, Andre Lukas, and Eirik E.\ Svanes}
\date{}
\maketitle
\begin{center}
\vskip -0.3cm
{\small {\it Rudolf Peierls Center for Theoretical Physics, Oxford University\;,\\
$~~~~~$ 1 Keble Road, Oxford, OX1 3NP, U.K.}\\}
\fnote{}{\hskip -0.6cm Michael.Klaput@physics.ox.ac.uk\\lukas@physics.ox.ac.uk\\Eirik.Svanes@physics.ox.ac.uk}

\end{center}
\vskip 0.8cm
\abstract{\noindent Compactifications of the heterotic string with NS flux normally require non Calabi-Yau internal spaces which are complex but no longer K\"ahler. We point out that this conclusion rests on the assumption of a maximally symmetric four-dimensional space-time and can be avoided if this assumption is relaxed. Specifically, it is shown that an internal Calabi-Yau manifold is consistent with the presence of NS flux provided four-dimensional space-time is taken to be a domain wall. These Calabi-Yau domain wall solutions can still be associated with a covariant four-dimensional $N=1$ supergravity. In this four-dimensional context, the domain wall arises as the ``simplest" solution to the effective supergravity due to the presence of a flux potential with a runaway direction. Our main message is that NS flux is a legitimate ingredient for moduli stabilization in heterotic Calabi-Yau models. Ultimately, the success of such models depends on the ability to stabilize the runaway direction and thereby ``lift" the domain wall to a maximally supersymmetric vacuum.
}
\thispagestyle{empty}
\newpage

\tableofcontents

\section{Introduction}
Ever since the mid 1980's when the standard embedding first was suggested \cite{Candelas198546}, the heterotic string compactified on Calabi-Yau manifolds has been an attractive way of obtaining the standard model from string theory. The appearance of the exceptional $E_8$ groups facilitates the construction of models with a realistic spectrum and, due to universality of the gauge couplings, gauge unification is natural in such constructions. Over the years, heterotic Calabi-Yau model building has been edging ever closer to realistic models and large classes of heterotic Calabi-Yau standard models have recently been constructed~\cite{Anderson:2012yf, Anderson:2011ns}.

Although heterotic Calabi-Yau models are attractive from the viewpoint of particle physics model building, moduli stabilisation has turned out to be more problematic than in type II theories. For one, this can be understood from the lack of RR fluxes in the heterotic string which leads to less flexibility in flux stabilization of moduli. However, even the use of NS flux seems problematic in the context of heterotic compactifications. It has been known since the work of Strominger~\cite{Strominger1986253} that the presence of NS flux in heterotic compactifications leads to internal manifolds which are complex but non-K\"ahler. This departure from Calabi-Yau manifolds, while not a problem in principle, constitutes a serious practical disadvantage. Compared to the significant body of knowledge on Calabi-Yau manifolds, not much is known about the required non-K\"ahler spaces and the construction of realistic particle physics models based on such spaces seems a long way off. So it appears that, at present, NS flux, the only type of flux available in heterotic models, is of little practical use in the context of realistic model building.

The main point of this paper is to explain how this conclusion can be avoided and to show that heterotic Calabi-Yau compactifications can indeed be consistent with the presence of NS flux. This requires dropping one of the assumptions which led to Strominger's result, namely that of maximal symmetry of the four-dimensional space-time. Instead, we will assume that four-dimensional space-time has the structure of a domain wall, with a $2+1$-dimensional maximally symmetric world-volume and one transverse direction. As we will show, any Calabi-Yau manifold and any harmonic NS flux on this manifold can be combined with such a domain wall to form a full 10-dimensional solution of the heterotic string, at least to lowest order in $\alpha'$.

At first sight, a non maximally symmetric four-dimensional space-time appears to be inconsistent with the usual requirements of homogeneity and isotropy of the universe. Let us discuss this point in more detail. First, note that the aforementioned 10-dimensional Calabi-Yau domain wall solutions can still be associated with a covariant, four-dimensional $N=1$ supergravity theory, obtained by a ``naive" compactification on the Calabi-Yau manifold with NS flux. This four-dimensional effective supergravity has a non-vanishing flux superpotential which fixes some of the moduli but still has runaway directions. For this reason, a maximally symmetric space-time does not solve this theory and the simplest solution is a half-BPS domain wall which corresponds to the four-dimensional part of the full 10-dimensional solution. From this point of view, the appearance of a non-maximally symmetric space-time simply indicates that not all moduli have been fixed and that the potential has runaway directions, a common occurrence in string models. Lifting the runaway direction - and, hence, lifting the domain wall to a maximally symmetric space - becomes a matter of moduli stabilization. Additional effects beyond the leading order solution, such as $\alpha'$ corrections, particularly due to the gauge bundle in heterotic theories~\cite{Anderson:2009sw,Anderson:2010mh}, and non-perturbative effects such as gaugino condensation can play a role in moduli stabilization and the ultimate fate of a solution rests on analysing the combination of all these effects. Analyzing moduli stabilization is not the main purpose of this paper. However, we note that it has recently been demonstrated~\cite{Klaput:2012vv}, in the somewhat different context of heterotic compactifications on coset spaces, that a combination of $\alpha'$ and non-perturbative corrections can indeed lift a four-dimensional domain wall to a maximally symmetric space-time. Our main point is to show that NS flux is a legitimate ingredient for moduli stabilization in the context of heterotic Calabi-Yau vacua. 

The paper is organised as follows. In section 2 we give a short recap of why Ricci-flat maximally symmetric compactifications do not allow for flux before we argue how non-maximally symmetric compactifications avoid this no-go result. In section 3 we specialise to domain wall compactifications and show that for every Calabi-Yau space there exists a domain wall solution with given harmonic NS flux. In section 4 we argue that the four-dimensional effective theories of regular Calabi-Yau vacua and Calabi-Yau domain wall vacua, differ essentially by the presence of a non-vanishing superpotential for the complex structure moduli. The proof that our constructions are valid away from the large complex structure limit in moduli space is given in the appendix. We conclude in section 5.

\section{Calabi-Yau compactifications and flux}
As a warm-up, we begin by reviewing the standard arguments for why NS flux is inconsistent with an internal Calabi-Yau manifolds, provided the four-dimensional space is maximally symmetric. It is then shown that these arguments break down if we allow the four-dimensional space-time to be a domain wall. The full 10-dimensional Calabi-Yau domain wall solution is presented in the next section.

\subsection{Maximally symmetric space-time}
We begin with the standard assumption that 10-dimensional space is a (possibly warped) product of a compact six-dimensional space $X_6$ and four-dimensional maximally symmetric space-time $M_4$ with metric
\begin{equation}\label{eq:ansatzmaxsymm}
ds^2_{10} = e^{2A(x^m)}\left(g_{\mu\nu}(x^\mu)\,\d x^\nu \otimes \d x^\nu + g_{mn}(x^m)\,\d x^m \otimes \d x^n\right)\; .
\end{equation}
Here $A$ is a warp factor, $g_{\mu\nu}$ with indices $\mu,\nu,\dots =0,1,2,3$ is a maximally symmetric metric on $M_4$ and $g_{mn}$ with indices $m,n,\dots =4,\ldots ,9$ an unspecified metric on $X_6$. As usual, we demand that some supersymmetry is unbroken by the compactification. The corresponding conditions, from the supersymmetry transformations of the gravitino and the dilatino, read
\begin{align}
\label{eq:killing1}
\Big(\nabla_M+\frac{1}{8}\mathcal{H}_M\Big)\epsilon=0\\
\label{eq:killing2}
\Big(\slashed\nabla\phi+\frac{1}{12}\mathcal{H}\Big)\epsilon=0\; .
\end{align}
Here, $M,N,\dots = 0,\ldots ,9$ are 10-dimensional indices, $\phi$ is the dilaton, and $\epsilon$ is a Majorana-Weil spinor parametrising supersymmetry. Further, we have defined the contractions $\mathcal{H}_M=H_{MNP}\Gamma^{NP}$ and $\mathcal{H}=H_{MNP}\Gamma^{MNP}$ of the NS three-form field, $H$.

The standard course of action is to set $H=0$ in those equations and study the resulting implications, leading to the well-known conclusion that $X_6$ must be a Calabi-Yau manifold with a Ricci-flat metric $g_{\mu\nu}$. Here, we are interested in the converse, namely assuming that $X_6$ is a Calabi-Yau manifold and analysing the implications for $H$. Any components of $H$ with four-dimensional indices must, of course, vanish due to four-dimensional maximal symmetry so we can focus on the purely internal components $H_{mnp}$. We begin by contracting eq.~(\ref{eq:killing1}) with $\Gamma^M$ and using eq.~(\ref{eq:killing2})  to get
\begin{equation}
\Big(\slashed\nabla-\frac{3}{2}\slashed\nabla\phi\Big)\epsilon=0,
\end{equation}
where the contractions are now over indices on the internal space $X_6$. For the re-scaled spinor $\tilde\epsilon=e^{-\frac{3}{2}\phi}\epsilon$ this implies $\slashed\nabla\tilde\epsilon=0$ and, hence, $\nabla_m\tilde\epsilon=0$. We conclude that $\tilde{\epsilon}$ is a covariantly constant spinor under the Levi-Civita connection. After a suitable $SO(6)$ redefinition of the gamma matrices we may assume that $\Gamma^a\tilde\epsilon=0$, where $a,b,\dots$ are holomorphic internal indices. Then, eq.~(\ref{eq:killing1}) leads to
\begin{equation}
\Big(3\nabla_m\phi+\frac{1}{4}\mathcal{H}_m\Big)\tilde\epsilon=0.
\end{equation}
Expanding in holomorphic and anti-holomorphic indices, and using $\{\Gamma^a,\Gamma^{\bar b}\}=2g^{a\bar b}$, this becomes
\begin{equation}
\Big(3\nabla_m\phi+\frac{1}{2}H_{ma\bar b}g^{a\bar b}+\frac{1}{4}H_{m\bar a\bar b}\Gamma^{\bar a\bar b}\Big)\tilde\epsilon=0.
\end{equation}
The last term implies $H_{m\bar a\bar b}=0$ and, since $H$ is  a real form, this leads to $H=0$. Then, it follows from the first term that $\nabla_m\phi=0$. Hence, we conclude that solving the supersymmetry conditions for a maximally symmetric four-dimensional space-time and an internal Calabi-Yau space, requires us to set $H=0$. This is the standard no-go theorem for flux on Calabi-Yau manifolds.

Is it possible to avoid this conclusion by relaxing the condition of unbroken supersymmetry? There is a simple argument~\cite{Gauntlett:2003cy} which shows that this does not change the situation, at least at zeroth order in $\alpha'$. To see this, let us recall that the dilaton equation of motion reads to zeroth order in $\alpha'$
\begin{equation}\label{eq:dilatoneom}
\nabla^2\,\e ^{-2\phi}=\e^{-2\phi} *(H\wedge * H)
\;,
\end{equation}
where $\nabla_M$ is the covariant derivative of the  Levi-Civita connection on $M_{10}$. With the ansatz \eqref{eq:ansatzmaxsymm} it becomes
\begin{equation}
-\d\left(\e^{4A}* \d \e ^{-2\phi}\right)=\e^{4A-2\phi} H\wedge * H
\;,
\end{equation}
where $d$ and $*$ denote the six-dimensional exterior derivative and Hodge star on $X_6$, respectively. Integrating over $X_6$ we obtain
\begin{equation}\label{eq:normHzero}
-\int_{X_6}\d_6\left(\e^{4A}*_6 \d_6 \e ^{-2\phi}\right)=\int_{X_6}\e^{4A}\e^{-2\phi} (H\wedge *_6 H)=\|\e^{2A}\e^{-\phi} H\|^2
\;.
\end{equation}
However, since $X_6$ is compact the integral on the left-hand side must vanish, which implies that $H=0$.

\subsection{Domain wall space-time}
Let us now modify the assumption of maximal symmetry and assume that four-dimensional space time is a domain wall, that is, $M_4=M_3\times Y$, where $M_3$ is a $2+1$-dimensional maximally symmetric space along the domain wall world volume and $Y$ is the single transverse direction. The corresponding 10-dimensional Ansatz now reads~\footnote{In general, $y$ and $x^m$ dependent warp factors can be introduced in the four-dimensional part of the metric. As we will see in the next section, these generalizations are not required for the class of solutions studied here and will, therefore, be omitted for simplicity.}
\begin{equation}\label{eq:ansatzdomainsusy}
ds^2_{10} = \eta_{\alpha\beta}(x^\alpha)\,\d x^\alpha \otimes \d x^\beta + \d y \otimes \d y + g_{mn}(x^m,y)\,\d x^m \otimes \d x^n\;,\quad
\phi=\phi(y,x^m)
\end{equation}
where $\eta_{\alpha\beta}$ with indices $\alpha,\beta,\dots =0,1,2$ is the Minkowski metric in three dimensions, $g_{mn}(x^m,y)$ is the metric on $X_6$ which is now fibred over the direction $y$, transverse to the domain wall. For simplicity, we will still require that all components of the flux $H$ with any indices in $M_4$ directions vanish. However, we note that the domain wall Ansatz is consistent with a more general structure of the flux~\cite{Gray:2012md}, specifically with non-zero components $H_{ymn}$ and $H_{\alpha\beta\gamma}\sim\epsilon_{\alpha\beta\gamma}$.

Let us briefly discuss how the no-go arguments presented in the previous sub-section are now being circumvented, beginning with the argument based on supersymmetry. Rather than leading to the condition $\slashed\nabla_6\tilde\epsilon=0$, the Killing spinor eqs.~\eqref{eq:killing1}, \eqref{eq:killing2} now imply that
\begin{equation}
\slashed\nabla_6\tilde\epsilon=\frac{3}{2}(\slashed\nabla_4\phi)\tilde\epsilon,
\end{equation}
where $\slashed\nabla_6$ and $\slashed\nabla_4$ denote the Dirac operators on the internal and external space respectively. Hence, for a $y$-dependent dilaton $\tilde\epsilon$ is no longer covariantly constant and the subsequent argument, leading to $H=0$, fails. 

The non-supersymmetric argument, based on the dilaton equation of motion, relied on the compactness of the internal space which led to the vanishing of the integral on the right-hand side of eq.~\eqref{eq:normHzero}. Its generalization to the domain wall case contains a seven-dimensional integral which can no longer be argued to vanish as it involves the non-compact $y$-direction. More precisely, the dilaton equation of motion \eqref{eq:dilatoneom} becomes
\begin{equation}
-\partial_y^2\e^{-2\phi}-\partial_y\e^{-2\phi}\, \partial_y(* 1)-\Delta_6\e^{-2\phi}=\e^{-2\phi} *(H\wedge * H)
\; ,
\end{equation}
where $*$ denotes the six-dimensional Hodge star. Dualising and integrating over $X_6$ we now obtain
\begin{equation}\label{eq:floweqdomain}
-\partial_y^2\e^{-2\phi}-\partial_y\e^{-2\phi}\, F(y)=\frac{1}{\mathrm{V}}\|\e^{-\phi} H\|^2
\;,
\end{equation}
where the function $F(y)$ describes the $y$-dependence of the volume form on $X_6$, $\partial_y(*_6 1)=F(y)\, *_6 1$, and $\mathrm{V}:= \int_{X_6}*_6 1$ is the volume of $X_6$. This is now a flow equation for $\phi(y)$ and it is possible to construct solutions for an arbitrary Calabi-Yau manifold $X_6$, as we will show in the next section. 

\section{Calabi-Yau domain walls and flux}\label{chap:cydomain10d}
In the previous section we saw that Calabi-Yau compactifications of the heterotic string with maximally symmetric four-dimensional space-time are inconsistent with the presence of flux. If we would like to add flux, we have to relax one of the underlying conditions. As motivated in the last section, we will relax the condition of four-dimensional maximal symmetry. Instead, we assume that four-dimensional space-time has the structure of a domain wall, $M_4=M_3\times Y$, with the associated 10-dimensional Ansatz~\eqref{eq:ansatzdomainsusy} for the metric and the dilaton and non-zero flux, $H_{mnp}$, on the internal space $X_6$ only. 

\subsection{Basic equations}
We require the compactification to preserve some supersymmetry. It is well-known that solutions of the supersymmetry conditions are also solutions to the equations of motion if they satisfy an additional integrability condition. This condition is automatically satisfied at zeroth order in $\alpha'$ as a consequence of the Bianchi identity \cite{Martelli:2010jx}.

Unbroken supersymmetry requires the internal space $X_6$ to have an $SU(3)$ structure which can be described by a two- and three-form $(J,\Omega)$. We denote the real and imaginary parts of $\Omega$ by $\Omega_\pm$. It can then be shown~\cite{Gauntlett:2002sc}, that the Killing spinor equations~\eqref{eq:killing1}, \eqref{eq:killing2} for this Ansatz are equivalent to
\begin{eqnarray}
 {\rm d}\Omega_-&=&2{\rm d}\phi\wedge \Omega_-
\label{eq1}
\\
 J\wedge {\rm d}J&=&J\wedge J\wedge {\rm d}\phi
\label{eq2}
\\
 J\wedge H&=& *{\rm d}\phi\label{eq3}
\\
\label{eq:killingspinor:original1}
{\rm  d}J&=&2\phi'\Omega_--\Omega'_--2{\rm d}\phi\wedge J+*H
\\
 {\rm d}\Omega_+&=&J\wedge J'-\phi'J\wedge J+2{\rm d}\phi\wedge\Omega_+
\\
 \Omega_-\wedge H&=&2\phi'* 1
\\
 \Omega_+\wedge H&=&0\; ,
\label{eq:killingOmegaplus}
 \end{eqnarray}
where primes denote $y$-derivatives and ${\rm d}$ and $*$ refer to the six-dimensional exterior derivative and Hodge dual on $X_6$, as before. These equations have to be supplemented with the $H$ equation of motion and the Bianchi identity (at zeroth order in $\alpha'$) which imply that $H$ must be $y$-independent and a harmonic three-form on $X_6$. 

What we would like to ask is, if the above system of equations can be solved for non-zero $H$, provided that $X_6$ is a Calabi-Yau manifold and $(J,\Omega)$ is the integrable $SU(3)$ structure with ${\rm d}J={\rm d}\Omega=0$. Then, the second eq.~\eqref{eq2} implies that the dilaton is a constant on $X_6$, ${\rm d}\phi=0$ and, as a result, the first three equations~\eqref{eq1}--\eqref{eq3} are satisfied. The remaining four equations specialize to the the flow-equations
\begin{align}\label{eq:killingspinor:1}
{\Omega'}_+&=2\phi '\Omega_+-H
\\\label{eq:killingspinor:2}
J\wedge J'&=\phi' J\wedge J\;
\\\label{eq:killingspinor:3}
\Omega_-\wedge H&=2\phi'\;*1\;,
\end{align}
and the constraint
\begin{align}
\label{eq:10dconstraints}
\Omega_+\wedge H =0 \;.
\end{align}
In \eqref{eq:killingspinor:1} we have used that the Hodge-dual and derivatives commute when acting on $\Omega$ explicitly, as is shown in appendix \ref{app:commute}. 

The equations~\eqref{eq:killingspinor:1}--\eqref{eq:killingspinor:3} are first-order differential equations which describe the variation of the $SU(3)$ structure $(J,\Omega)$ and the dilaton $\phi$ along the $y$-direction. By expanding $J$ and $\Omega$ into a basis of harmonic two- and three-forms they can be broken up into a set of scalar first-order differential equations whose solutions exist from general theorems. Further, for fixed flux $H$, eq.~\eqref{eq:10dconstraints} represents a constraint on the complex structure. Let us now analyse this in more detail.

\subsection{Existence of solutions}
The existence of solutions to the above system of equations, \ eqs. \eqref{eq1}--\eqref{eq:killingOmegaplus}, has been established in general, see for example ref.~\cite{Hitchin2001}. However, it is useful to construct a solution explicitly, in order to gain more detailed insight, in particular about its asymptotic behaviour at large distance from the domain wall.

To analyse eqs.\ \eqref{eq:killingspinor:1}--\eqref{eq:10dconstraints} in detail, we introduce a symplectic basis $\{\alpha_A,\beta^B\}$ of harmonic three forms and a basis $\{\omega_i\}$ of harmonic two forms on $X_6$. As usual, the $SU(3)$ structure forms $(J,\Omega)$ can then be expanded as
\begin{equation}\label{eq:expOmZ}
 J=v^i\omega_i\; ,\qquad \Omega=Z^A\alpha_A-\mathcal{G}_{B}\beta^B\; ,
\end{equation} 
where $v^i$ and $Z^A$ are the K\"ahler and complex structure moduli, respectively, and the functions $\mathcal{G}_{B}$ are the first derivatives of the pre-potential ${\mathcal G}={\mathcal G}(Z)$. We also introduce the volume
\begin{equation}
 V=\frac{1}{6}\int_{X_6} J\wedge J \wedge J=\frac{1}{6}d_{ijk}v^iv^jv^k\; , \label{Vdef}
\end{equation} 
with the triple intersection numbers $d_{ijk}$. For more details on the description of the Calabi-Yau moduli space, see appendix \ref{app:intro}. Likewise, the expansion of the flux in terms of the symplectic basis can be written as
\begin{equation}
\label{eq:fluxexp}
H =\mu^A\alpha_A+\epsilon_B\beta^B\; ,
\end{equation}
where $\mu^A$ and $\epsilon_A$ are the flux parameters. It is useful to introduce the re-scaled complex structure moduli $X^A=e^{-2\phi}Z^A$. Since the functions ${\mathcal G}_A$ are homogeneous of degree one it follows that ${\mathcal G}_A(X)=e^{-2\phi}{\mathcal G}_A(Z)$. We also define a new coordinate $z$ by 
\begin{equation}
\label{eq:variablechange}
 \frac{dy}{dz}=e^{2\phi}\; .
\end{equation} 
Using the above expansions and definitions, the flow equations~\eqref{eq:killingspinor:1}--\eqref{eq:killingspinor:3} can be re-written in the form
\begin{eqnarray}
\label{eq:killingmod1}
 \partial_z\,{\rm Re}(X^A)&=&-\mu^A\\
\label{eq:killingmod2}
 \partial_z\,{\rm Re}({\mathcal G}_A)&=&\epsilon_A\\
 \partial_z\, v^i&=&\partial_z\phi\\
 \partial_z\phi&=&\frac{e^{4\phi}}{2V}\textrm{Im}\left(\epsilon_AZ^A + \mu^A\mathcal{G}_A \right)\label{dileq}
\end{eqnarray} 
while the constraint \eqref{eq:10dconstraints} takes the form
\begin{equation}
 \textrm{Re}\left(\epsilon_AX^A +  \mu^A\mathcal{G}_A\right)=0\; . \label{cons}
\end{equation} 
Here and in the following $\mathcal{G}_A$ should be interpreted as functions of the re-scaled complex structure moduli $X^A$. The first three of these equations are easily integrated leading to
\begin{equation}
 \textrm{Re}(X^A)=-\mu^Az-\gamma^A\;,\quad
 \textrm{Re}(\mathcal{G}_B)=\epsilon_Bz+\eta_B\;,\quad
 v^i=e^\phi v^i_0\; , \label{vsol}
\end{equation} 
where $\gamma^A$, $\eta_B$ and $v_0^i$ are integration constants.  For a given Calabi-Yau three-fold, the $\mathcal{G}_A$ are known (although complicated) functions of the complex structure moduli $X^A$.  Hence, the above equations implicitly determine the $z$-dependence of $X^A$. With these solutions eq.~\eqref{cons} turns into
\begin{equation}
\label{eq:cons}
-\gamma^A\epsilon_A+\eta_B\mu^B=0\;,
\end{equation}
that is, a condition on the integration constant which can be satisfied by a suitable choice of these constants\footnote{In ref.~\cite{Lukas:2010mf}, further constraints for the existence of a solution, in addition to \eqref{eq:cons}, are given. These arise due to the assumption that the flux components $\{\epsilon_0,\mu^0\}$ vanish, which is required for the half-flat compactifications discussed in ref.~\cite{Lukas:2010mf} but can be avoided for the Calabi-Yau compactifications discussed here.}. Finally, we need to discuss the dilaton equation~\eqref{dileq}. First, we note that, from eqs.~\eqref{vsol} and \eqref{Vdef}, the volume is given by $V=V_0e^{3\phi}$, where $V_0$ is a constant explicitly given by $V_0=d_{ijk}v_0^iv_0^jv_0^k/6$. Inserting this into the dilaton flow equation~\eqref{dileq} we obtain
\begin{equation}
\label{eq:eqdilaton}
\partial_ze^{-\phi}=\frac{1}{2V_0}(\epsilon_A\,\textrm{Im}\;X^A+\mu^B\,\textrm{Im}\;\mathcal{G}_B)\; .
\end{equation}
With the explicit solutions for $X^A$ this leads to an explicit, although complicated first order differential equation for the $z$-dependence of the dilaton which can, at least in principle, be integrated.

In summary, we have established the existence of supersymmetric domain wall solutions for any choice of Calabi-Yau manifold and any harmonic flux on it.

\subsection{Asymptotic behaviour of solutions}
Existence of solutions to the flow-equations \eqref{eq:killingspinor:1}--\eqref{eq:killingspinor:3} is always guaranteed as we have demonstrated above. However, explicit integration requires detailed knowledge of the pre-potential and can only be done on a case-by-case basis. Still, we can deduce the properties of the solution in the limit of large $y$, that is, the behaviour of the fields $\{\phi,X^A, v^i\}$ far away from the domain wall.

To do this, we return to the flow equations \eqref{eq:killingspinor:1}--\eqref{eq:killingspinor:3} for a moment. Recall that equation \eqref{eq:killingspinor:1} is equivalent to
\begin{equation}
\p_y(e^{-2\phi}\Omega_-)=e^{-2\phi}*H.
\end{equation}
Multiplying \eqref{eq:killingspinor:3} with $e^{-2\phi}$ and applying $\p_y$, we get
\begin{equation}
\label{eq:ddphi}
\p_y^2(e^\phi)=-\frac{1}{2e^{2\phi}V_0}\int_XH\wedge*H=-\frac{1}{2e^{2\phi}V_0}\vert\vert H\vert\vert^2,
\end{equation}
where we also have integrated over $X$. Note that \eqref{eq:ddphi} implies that the strictly positive function $e^\phi$ has a negative second order derivative. Its derivative must, therefore, approach some non-negative constant from above. In fact, for non-vanishing flux this constant cannot be zero since eq.~\eqref{eq:ddphi} would then imply
\begin{equation}
\lim_{y\rightarrow\infty}\vert\vert H\vert\vert^2=0\;.
\end{equation}
Since the flux is $y$-independent this can only be true if $H=0$. Hence, with non-vanishing flux on $X$, the dilaton $e^\phi$ approaches a linear increasing function as $y\rightarrow\infty$. The generic $y$-dependence of $e^\phi$ and its derivative has been plotted below.
\begin{figure}[h!]
\begin{center}
\includegraphics[height=40mm]{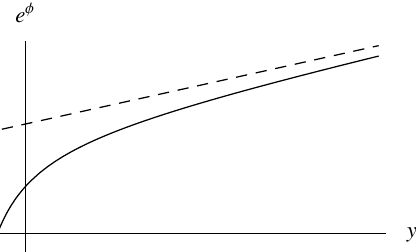}\;\;\;\;\;\;\;\;\;\;\;\;
\includegraphics[height=40mm]{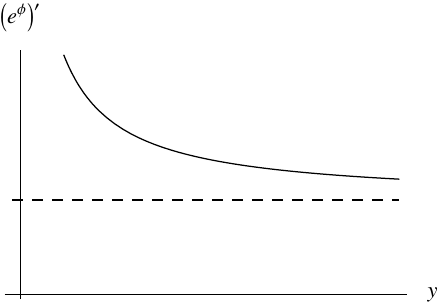}
\label{fig:asymptote}
\caption{\it Plot of the generic asymptotic behaviour of $e^\phi$ and its derivative $(e^\phi)'$ as $y\rightarrow\infty$.}
\end{center}
\end{figure}

Furthermore, from the definition of $z$ in eq.~\eqref{eq:variablechange} we see that in the limit $y\rightarrow\infty$ the behaviour of $e ^\phi$ implies that $z$ approaches a constant as $y\rightarrow\infty$. Accordingly, it follows that the rescaled fields $X^A$ approach constant values, while the original moduli $Z^A$ diverge. This means that the solution approaches the large complex structure limit far away form the domain wall, where the pre-potential can generically be approximated by
\begin{equation}
\mathcal{G}(Z)=\mathcal{K}_{ABC}Z^AZ^BZ^C
\;,
\end{equation}
with intersection numbers $\mathcal{K}_{ABC}$. This observation allows us to check consistency with the results obtained in ref.~\cite{Lukas:2010mf}. Indeed, inserting the above form of the pre-potential into eqs.~\eqref{eq:killingmod1}--\eqref{cons} and, in addition, setting $\epsilon_0=\mu^0=0$, yields precisely the solution given in ref.~\cite{Lukas:2010mf}.

\section{Four-dimensional low energy theory}\label{chap:lowenergy}
We will now discuss the effective four-dimensional theories associated to Calabi-Yau domain wall solutions. This four-dimensional theories are covariant, $N=1$ supergravities, identical to the ones obtained from compactification on Calabi-Yau manifolds without flux, apart from the presence of a non-vanishing superpotential for the complex structure moduli. The domain wall can be recovered as a BPS solution of the four-dimensional theory which couples to this superpotential.

We begin by reviewing the structure of the four-dimensional effective theory and its domain wall solution. Then we discuss the matching of the 10-dimensional Calabi-Yau domain wall solution, introduced in section 3, with the four-dimensional domain wall solution. We leave any technicalities for appendix \ref{app:comparison} as they would distract from the main point of the section. Our discussion extends the results of ref.~\cite{Lukas:2010mf} where matching has been shown in the large complex structure limit. Here we find that these results can be extended to the entire moduli space. We also comment on the asymptotic behaviour found in the previous section, now from a four-dimensional perspective. 

\subsection{$4d$ effective theory}\label{sec:4deff}
Upon dimensional reduction of the heterotic supergravity with a Calabi-Yau internal space $X_6$ one obtains a four-dimensional, $\mathcal{N}=1$ supergravity theory. It contains a set of chiral superfields $\Phi^X=(S,T^i,X^A)$ which correspond to the axio-dilaton $S=a+\I \,e^{-2\phi_4}$, the K\"ahler moduli $T^i$ and complex structure moduli $X^A$. Their kinetic terms are derived from the K\"ahler potential
\begin{align}\label{eq:kaehler4d}
K(\Phi^X,{\bar \Phi} ^ {\bar X})
=
K^{S}+K^{T}+K^{X}
=
- \ln \I ({\bar S}-S)
- \ln 8 V
 - \ln \I \Big[\mathcal{G}_B\bar X^B-X^A\mathcal{\bar G}_A\Big]
\;.
\end{align}
Here $V$ corresponds to the volume of $X_6$ and can be expressed in terms of its intersection numbers $d_{ijk}$, i.e.\ $V=\frac{1}{6}d_{ijk}t^i t^j t^k$ with $t^i = {\rm Im}\, T^i$. 

The superpotential of the theory is now given by
\begin{equation}
\label{eq:superpotential}
W=\sqrt{8}(\epsilon_A X^A+\mu^A\mathcal{G}_A),
\end{equation}
where $\mu^A, \epsilon_A$ are the flux parameters as defined in eq.~\eqref{eq:fluxexp}, $\mathcal{G}=\mathcal{G}(X^A)$ is the prepotential for the complex structure moduli and $\mathcal{G}_A=\frac{\partial}{\partial X^A}\mathcal{G}$ its derivatives. Recall that $\mathcal{G}$ is a homogeneous function of degree two. 

\subsection{Domain wall solution}
It has been shown \cite{Lukas:2010mf} that the four-dimensional theories just described have $1/2$ BPS domain wall solutions with metric
\begin{align}\label{eq:4ddomainwallmetric}
\mathrm{d}s^2=\e^{-2\phi(y)}\left(\eta_{\alpha\beta}\mathrm{d}x^\alpha\mathrm{d}x^\beta + \mathrm{d}y^2\right)
\end{align}
where $\eta_{\alpha\beta}\mathrm{d}x^\alpha\mathrm{d}x^\beta $ is the 1+2 dimensional Minkowski metric and $y=x^3$. With this metric the Killing-spinor equations reduce to
\begin{align}\label{eq:4dKS}
\p_{y} \Phi^X=-ie^{-\phi_4}e^{K/2}K^{X\bar Y}D_{\bar Y}{\bar W}
\;,
\end{align}
together with the constraint that the superpotential $W$ has to be purely imaginary and the axionic components of all fields are constant. Here, $D_{ Y}W = \partial_Y W + K_Y W$ as usual.

Furthermore, it was shown in ref.~\cite{Lukas:2010mf} that such four-dimensional domain wall solutions match their 10-dimensional counterparts, discussed in section~\ref{chap:cydomain10d}. However, the matching in ref.~\cite{Lukas:2010mf} was carried out only in the large complex structure limit, as is appropriate for the half-flat compactifications discussed there. 
For Calabi-Yau manifolds a restriction to large large complex structure is unnecessary. Fortunately, it turns out that this requirement is merely technical and that the matching can be shown to hold everywhere in complex structure moduli space. In detail, this is proven in appendix \ref{app:comparison}, but we briefly review the procedure and results here.

\subsection{Comparing four- and 10-dimensional solutions}

The matching between the four- and 10-dimensional domain-wall vacua is carried out by showing that the respective Killing spinor equations are equivalent under appropriate field re-definitions. We merely present the results here, while the full proof is given in appendix \ref{app:comparison}.

It turns out that for the matching to work we need to relate four- and 10-dimensional fields as
\begin{align}
\label{eq:10d4d:redef:1}
\e^{2\phi}&=\e^{2\phi_4}V/V_0
\\\label{eq:10d4d:redef:2}
Z^A&=\e^{2\phi}X^A
\\\label{eq:10d4d:redef:3}
v^i&=t^i
\;.
\end{align}
This demonstrates that the low energy description of heterotic domain walls are given by the domain wall solutions of the $N=1$ four-dimensional supergravity theories discussed in the previous section. The matching holds everywhere in complex structure moduli space and for general harmonic flux.

We also comment briefly on the large $y$-behaviour we found in the previous section. We saw that the fields $X^A$ stabilise at constant values as $y\rightarrow\infty$ far away from the domain-wall, where its influence is negligible. This is expected from a four-dimensional point of view, as we have introduced a superpotential for these fields. No such superpotential has been introduced for the dilaton or the K\"ahler moduli, which remain unstabilised.

As for the four-dimensional dilaton, we saw in the last section that the 10-dimensional dilaton diverges as $y\rightarrow\infty$. From \eqref{eq:10d4d:redef:1} and the fact that $V=e^{3\phi}V_0$, we see that
\begin{equation}
\p_y\phi=-2\p_y\phi_4\; .
\end{equation}
Hence, the four-dimensional dilaton goes to negative infinity, and we thus approach the weak coupling regime far away from the domain wall.

\section{Conclusions}
In this paper, we have shown that heterotic Calabi-Yau compactifications with flux exist, provided that we relax the condition of having a maximally symmetric four-dimensional space-time. Using a four-dimensional domain-wall ansatz instead, we have found Calabi-Yau domain wall solutions for any harmonic flux and throughout complex structure moduli space. This extends previous results obtained in the large complex structure limit.

The main message is that harmonic NS flux is a legitimate ingredient in heterotic Calabi-Yau compactifications and can be added to the model without deforming the Calabi-Yau to a non-Kahler manifold. This means that the powerful set of model-building tools on Calabi-Yau manifolds is available while NS flux can be added as a useful ingredient for moduli stabilization.

Ultimately, the success of these models depends on the ability to lift these domain wall vacua to maximally symmetric ones which amounts to stabilizing the remaining moduli, that is, the dilaton and the T-moduli. In ref.~\cite{Klaput:2012vv}, it has been shown that this can indeed be achieved in certain half-flat domain wall compactifications based on group coset spaces. Whether these results carry over to the present Calabi-Yau domain wall solutions is a subject of future study.

Another obvious generalization of the present work is to search for 10-dimensional heterotic solutions based on Calabi-Yau manifolds, harmonic flux and more general four-dimensional BPS solutions, including, for example, four-dimensional cosmic string and black hole solutions. Especially, Calabi-Yau black hole solutions might be interesting in this context, as they might turn out to be consistent with the present universe without the need to ``lift" to a maximally symmetric four-dimensional space-time. Work in this direction is currently underway.

\section*{Acknowledgements}
We would like to thank James Gray for useful discussions. A.~L.~is supported in part by the EPSRC network grant EP/I02784X/1. M.~K.\ is supported by an STFC scholarship. E.~E.~S. is supported by the Clarendon Scholarship and the Balliol Dervorguilla Scholarship.

\appendix
\section{Calabi-Yau symplectic geometry}\label{app:intro}
Here, we briefly summarize some useful facts about the symplectic geometry of Calabi-Yau moduli spaces~\cite{Candelas:1990pi}. Overviews may be found, for example, in refs.~\cite{CyrilThesis, MatthiasThesis, Ceresole:1995ca}.

\subsection{Harmonic expansion}
The K\"ahler form $J$ is expanded in a basis of harmonic $(1,1)$-forms
\begin{equation}
J=v^i\omega_i,
\end{equation}
where $\omega_i\in H^{(1,1)}(X)$. Likewise the harmonic (3,0)-form $\Omega$ is expanded as
\begin{equation}
\Omega=Z^A\alpha_A-\mathcal{G}_B\beta^B,
\end{equation}
where $\{\alpha_A,\beta^B\}\in H^3(X)$ is a real symplectic basis such that
\begin{equation}
\int_X\alpha_A\wedge\beta^B=\delta_A^B,
\qquad
\int_X\alpha_A\wedge\alpha_B=\int_X\beta^A\wedge\beta^B=0.
\end{equation}
The complex structure moduli space is a K\"ahler manifold, described by a holomorphic pre-potential $\mathcal{G}=\mathcal{G}(Z)$ which is a homogeneous function of degree two. Its derivatives are denoted by $\mathcal{G}_A=\p_A\mathcal{G}=\frac{\p\mathcal{G}}{\p Z^A}$.

Note that, in the context of the Calabi-Yau domain walls we discuss, the $\SU3$ structure forms $J$ and $\Omega$ will depend on $y$, the direction normal to the domain wall. However, the basis forms $\{\omega_i\}$ and $\{\alpha_A,\beta^B\}$ are related to cycles of the Calabi-Yau manifold and are, hence, independent of $y$. Consequently, the $y$-dependence entirely resides in the moduli-fields $\{v^i,Z^A\}$.

\subsection{Some symplectic geometry and the Hodge star}
We adopt the convention
\begin{equation}
*\Omega=-i\Omega.
\end{equation}
It then follows that\footnote{Note the different convention for the Hodge-star in \cite{Ceresole:1995ca}, which takes the complex conjugate after taking the dual.}
\begin{equation}
\label{eq:dualdOm}
*\p_A\Omega=i\p_A\Omega.
\end{equation}
The Hodge stars of the symplectic basis $\{\alpha_A,\beta^B\}$ are given by
\begin{align}
*\alpha_A&={A_A}^B\alpha_B+B_{AB}\beta^B\\
*\beta^A&=C^{AB}\alpha_B+{D^A}_B\beta^B,
\end{align}
where ${A_A}^B=-{D^A}_B$. These matrices may be written in terms of the matrix $N_{AB}$ given by
\begin{equation}
N_{AB}=\mathcal{\bar G}_{AB}+2i\frac{\textrm{Im}(\mathcal{G}_{AC})Z^C\textrm{Im}(\mathcal{G}_{BD})Z^D}{\textrm{Im}(\mathcal{G}_{CD})Z^CZ^D}.
\end{equation}
The corresponding expressions are
\begin{align}
\label{eq:A}
A&=(\textrm{Re}N)(\textrm{Im}N)^{-1}\\
\label{eq:B}
B&=-(\textrm{Im}N)-(\textrm{Re}N)(\textrm{Im}N)^{-1}(\textrm{Re}N)\\
\label{eq:C}
C&=(\textrm{Im}N)^{-1}.
\end{align}

\subsection{Useful identities}
Next, we give some identities which will be useful in the next sections. We first define the complex structure K\"ahler potential
\begin{equation}
\label{eq:cplxK}
\mathcal{K}=\ln\I\int\Omega\wedge\bar{\Omega}.
\end{equation}
Next we define the parameters
\begin{equation}
\label{eq:fs}
f_A^B=\p_AZ^B+K_AZ^B=D_AZ^B.
\end{equation}
It may then be shown that the matrix $N$, the K\"ahler potential, the parameters $f_A^B$ and the pre-potential satisfy the following identities
\begin{align}
\label{eq:usefulid1}
\mathcal{K}_{\bar BC}&=-\frac{1}{4V}(\textrm{Im}N)_{DE}\bar f^{\bar D}_{\bar B}f^E_C\\
\label{eq:usefulid3}
\bar N_{\bar A\bar B}f^B_C&=\mathcal{G}_{AB}f^B_C\\
\label{eq:usefulid2}
(\textrm{Im}N_{AB})\bar f^{\bar A}_{\bar C}\bar Z^{\bar B}&=0.
\end{align}

\section{Hodge star and y-derivatives}\label{app:commute}

In this appendix, we wish to prove that
\begin{align}
\p_y*\Omega=*\p_y\Omega
\;,
\end{align}
where $*$ denotes the six-dimensional Hodge dual on the Calabi-Yau manifold. This relation will be useful for proving the results in appendix \ref{app:comparison}.

First, from the harmonic expansion \eqref{eq:expOmZ} it follows that $\p_A\Omega=\alpha_A-\mathcal{G}_{AB}\beta^B$,  where we used the fact that $\mathcal{G}_{AB}$ is a homogeneous polynomial of degree zero, i.e. 
\begin{equation}
Z^A\delta\mathcal{G}_{AB}=Z^A\mathcal{G}_{ABC}\delta Z^C=0, 
\end{equation}
as $Z^A\mathcal{G}_{ABC}=0$. However, by imaginary self duality of $\Omega$ we have
\begin{equation}
*\Omega=i Z^A(\alpha_A-\mathcal{G}_{AB}\beta^B)
\;.
\end{equation}
Using \eqref{eq:dualdOm}, this means that
\begin{align}
*\delta\Omega&=\delta Z^A*(\alpha_A-\mathcal{G}_{AB}\beta^B)\\
&=\delta Z^Ai(\alpha_A-\mathcal{G}_{AB}\beta^B)\\
&=\delta\Big(Z^Ai(\alpha_A-\mathcal{G}_{AB}\beta^B)\Big)\\
&=\delta*\Omega,
\end{align}
for any first-order variation $\delta$. As before, we have used that $\mathcal{G}_{AB}$ is homogeneous of degree zero. Hence the Hodge dual and derivatives commute when acting on $\Omega$. 

\section{Matching 10- and four-dimensional equations}
\label{app:comparison}
In this appendix, we would like to show that the Killing-Spinor equations in 10 and four dimensions match, under a suitable field re-definition. 

Let us start by clearly stating the field redefinitions which will be necessary to relate both solutions. The dilaton $\phi$, K\"ahler moduli $v^i$ and complex structure moduli $Z^A$ of the 10-dimensional theory are related to the respective fields, $\phi_4$, $t^i$, $X^A$ of the four-dimensional theory via
\begin{align}
\label{eq:10d4d:redefinitions:1}
\e^{2\phi}&=\e^{2\phi_4}V/V_0
\\\label{eq:10d4d:redefinitions:2}
Z^A&=\e^{2\phi}X^A
\\\label{eq:10d4d:redefinitions:3}
v^i&=t^i
\;,
\end{align}
where again $V$ is the volume of the Calabi-Yau manifold $X_6$ and $V_0$ is some fixed reference volume. With these identifications, the equations for the K\"ahler moduli \eqref{eq:killingspinor:2} and \eqref{eq:4dKS} can be easily confirmed to match, in complete analogy to the proof in ref.~\cite{Lukas:2010mf}. 

Let us now demonstrate the matching of the Killing spinor equations for the dilaton whose four-dimensional version \eqref{eq:4dKS} becomes
\begin{align}\label{eq:dilatonflow4d:generic1}
\p_y\phi_4=\frac{\I\,\e^{2\phi_4}}{4}W\; .
\end{align}
Here, we have used the K\"ahler potential and superpotential from eqs.~\eqref{eq:kaehler4d} and \eqref{eq:superpotential}. From the relation \eqref{eq:10d4d:redefinitions:1} between the 10- and four-dimensional dilaton, and the $y$-dependence of the volume 
\begin{equation}
\p_yV=3\p_y\phi V\; ,
\end{equation}
implied by eq.~\eqref{eq:killingspinor:2}, it follows that $\partial_y\phi=-2\partial_y\phi_4$. With the last relation it can be easily seen that \eqref{eq:dilatonflow4d:generic1} matches the 10-dimensional  dilaton equation \eqref{eq:killingspinor:3}, upon integrating the latter equation over $X$.

Next, let us show the matching of the Killing spinor equations for the complex structure moduli. To see this, we start with the 10-dimensional equation~\eqref{eq:killingspinor:original1}, given by
\begin{align}
\p_y\Omega_-=2(\p_y\phi)\Omega_-+*H\;.
\end{align}
This can be turned into an equation for $\Omega$ if we commute the Hodge star and $y$-derivative, as shown to be valid in appendix \ref{app:commute}. This leads to
\begin{align}\label{eq:OmegaH}
\p_y\Omega=2(\p_y\phi)\,\Omega - (H - i\, *H)
\;.
\end{align}
If we expand $\Omega$ and $H$ with respect to a symplectic basis $(\alpha_A,\beta^A)$ as before, that is,
\begin{align}\label{eq:omegaHexpansions}
\Omega=Z^A\left(\alpha_A - \mathcal{G}_{AB}\beta^B\right)
\; ,\qquad
H=\mu^A\alpha_A + \epsilon_A\beta^A
\; ,
\end{align}
we can turn \eqref{eq:OmegaH} into an the equation
\begin{equation}
\p_y(e^{-2\phi}Z^A)=-e^{-2\phi}(\mu^A-i\tilde\mu^A).
\label{eq:10dflow}
\end{equation}
for the complex structure moduli $Z^A$.  Here $\tilde\mu^A=C^{AB}\epsilon_B+A_B^A\mu^B$. With the complex structure K\"ahler potential \eqref{eq:cplxK}, equation \eqref{eq:10dflow} can be written in terms of complex structure moduli space geometry as 
\begin{align}
\p_y(e^{-2\phi}Z^A)
&=
-e^{-2\phi}\mathcal{K}^{A\bar B}\mathcal{K}_{\bar B C}C^{CB}\Big((C^{-1})_{BD}\mu^D
-i\epsilon_B-i(\textrm{Re}N)_{BD}\mu^D\Big)
\\
&=
-i\frac{e^{-2\phi}}{4V}\mathcal{K}^{A\bar B}(\textrm{Im}N)_{DE}
\bar f^{\bar D}_{\bar B}
f^E_CC^{CB}
\Big(
i(C^{-1})_{BD}\mu^D+\epsilon_B+(\textrm{Re}N)_{BD}\mu^D
\Big),
\end{align}
where the first equality follows from ${A_C}^A=C^{AB}(\textrm{Re}N)_{BC}$, and for the second equality we have used equation of \eqref{eq:usefulid1}. Using \eqref{eq:usefulid2}, and the fact that $C=(\textrm{Im}N)^{-1}$, we see that
\begin{align}
\label{eq:Killing10}
\p_y(e^{-2\phi}Z^A)&=-\frac{ie^{-2\phi}}{4V}\mathcal{K}^{A\bar B}\bar f^{\bar C}_{\bar B}\Big(\epsilon_C+N_{CD}\mu^D\Big)\notag\\
&=-\frac{ie^{-2\phi}}{4V}\mathcal{K}^{A\bar B}\bar f^{\bar C}_{\bar B}\Big(\epsilon_C+\mathcal{\bar G}_{\bar C\bar D}\mu^D\Big),
\end{align}
where in the last equality we have used \eqref{eq:usefulid3}.

We want to compare this the to the $4d$ Killing spinor equation \eqref{eq:4dKS} for the moduli $X^A$, which reads 
\begin{align}\label{eq:4dKSeq}
\p_{y}X^A=-\frac{i}{4} \e^{2\phi_4} K^{A\bar B}D_{\bar{B}}\bar{W}
=
-\frac{i}{4} \e^{2\phi_4} K^{A\bar B}D_{\bar{B}}X^{\bar{C}}(\epsilon_{\bar C}+\mathcal{\bar G}_{\bar C\bar D}\mu^{\bar D})
\;.
\end{align}
If we now use \eqref{eq:10d4d:redefinitions:2} to re-express all $\partial/\partial X^A$ derivatives into $\partial/\partial Z^A$ and the fact that $f_A^B=\p_A Z^B+\p_A \mathcal{K} Z^B=D_A Z^B$, then we see that in fact \eqref{eq:Killing10} and \eqref{eq:4dKSeq} are equal.

It remains to be shown that the constraint \eqref{eq:10dconstraints} is satisfied by the four-dimensional solution. To see this, recall that the four-dimensional Killing spinor equations force the superpotential to be purely imaginary. In 10 dimensions the analogous requirement is given by eq.~\eqref{eq:omegaHexpansions} which translates into a purely imaginary superpotential via the Gukov-Vafa-Witten formula
\begin{align}
W\propto\int_{X_6}H\wedge\Omega\; .
\end{align}

\providecommand{\href}[2]{#2}\begingroup\raggedright\endgroup

\end{document}